\documentclass[journal]{IEEEtran}
\usepackage{graphicx}
\usepackage{latexsym}
\usepackage{amsthm}
\usepackage{amsfonts}
\usepackage{subfigure}
\usepackage{dsfont}
\usepackage{epstopdf}
\usepackage{threeparttable}
\usepackage{multirow}
\usepackage{enumerate}
\usepackage{epsfig}
\usepackage{booktabs}
\usepackage{bm}
\usepackage{cite}
\usepackage{textcomp}
\usepackage{booktabs}
\usepackage{tabu}
\usepackage{color}
\usepackage{amsmath}
\usepackage{algorithm}
\usepackage{algpseudocode}
\usepackage{tabu}
\usepackage{booktabs}
\usepackage{diagbox}
\usepackage{bbding}
\newtheorem{remark}{Remark}

\usepackage{makecell}
\def\BibTeX{{\rm B\kern-.05em{\sc i\kern-.025em b}\kern-.08em
T\kern-.1667em\lower.7ex\hbox{E}\kern-.125emX}}
\begin{document}
\title{ELM-based Timing Synchronization for OFDM Systems by Exploiting Computer-aided\\Training Strategy}

\author{Mintao~Zhang,~Shuhai~Tang,~Chaojin~Qing,~Na~Yang,~Xi~Cai,~and~Jiafan~Wang
\thanks{The authors (\emph{Corresponding author: Chaojin Qing}) are with the School of Electrical Engineering and Electronic Information, Xihua University, Chengdu  610039, China (E-mails: zhangmt@mail.xhu.edu.cn, tangshh@stu.xhu.edu.cn, qingchj@mail.xhu.edu.cn, yangna6717@163.com, caixi-1983@163.com, and jifanw@gmail.com).}
}

\maketitle

\begin{abstract}
    Due to the implementation bottleneck of training data collection in realistic wireless communications systems, supervised learning-based timing synchronization (TS) is challenged by the incompleteness of training data. To tackle this bottleneck, we extend the computer-aided approach, with which the local device can generate the training data instead of generating learning labels from the received samples collected in realistic systems, and then construct an extreme learning machine (ELM)-based TS network in orthogonal frequency division multiplexing (OFDM) systems. Specifically, by leveraging the rough information of channel impulse responses (CIRs), i.e., root-mean-square (r.m.s) delay, we propose the loose constraint-based and flexible constraint-based training strategies for the learning-label design against the maximum multi-path delay. The underlying mechanism is to improve the completeness of multi-path delays that may appear in the realistic wireless channels and thus increase the statistical efficiency of the designed TS learner. By this means, the proposed ELM-based TS network can alleviate the degradation of generalization performance. Numerical results reveal the robustness and generalization of the proposed scheme against varying parameters.
\end{abstract}

\IEEEpeerreviewmaketitle
\section{Introduction}\label{I:I}
Orthogonal frequency division multiplexing (OFDM), as a kind of the modern wireless and mobile communication systems, has been subject to extensive research efforts not only from the 5G field communication systems \cite{ref:5Gfiled} but also from the wireless local area networks (WLANs) \cite{ref:Wifi} and the Internet-of-Things (IoT) systems \cite{ref:IoT}.
For the OFDM systems, timing synchronization (TS) plays a pivotal role in overall system performance, which aims to locate starting of the receiver fast Fourier transform (FFT) window within the inter-symbol-interference (ISI)-free region\cite{ref:ISIf3}.
For the TS, extensive studies have been conducted so far to improve the timing metric for OFDM systems \cite{ref:SC,ref:Minn,ref:park1,ref:CTS,ref:TSZC,ref:ZCTS}. However, classic TS schemes inevitably encounter nonlinear effects, e.g., channel fading, hardware imperfections, etc, posing a great challenge to avoid the ISI in the TS stage, and affecting the subsequent signal processing, e.g., channel estimation \cite{ref:JSCE} and symbol detection \cite{ref:JTSDe}, etc.

Due to its powerful capability in coping with nonlinear problems \cite{ref:LY1}, machine learning (ML) is a promising solution to tackle the challenge of nonlinear effects in the TS stage and also aroused extensive interest in the design of the physical layer communications \cite{ref:MLCui,ref:Cui2,ref:NDL3,ref:DL-CE2,ref:PC,ref:DL3} over the past few years.
In ML-based applications, substantial researches have proven that the predictive performance of supervised learning outperforms that of unsupervised learning because the supervised ones could directly take feedback for the prediction \cite{ref:SupL}.
As for synchronization, several supervised learning-based schemes~\cite{ref:CSIlearn,ref:DLsys,ref:DLE2Esync,ref:DL-PDCFO,ref:ELM-FTS} have been investigated.
For example in \cite{ref:CSIlearn}, a deep neural network (DNN)-based estimator was constructed to estimate the timing offset and carrier frequency offset (CFO).
On the basis of DNN-based auto-encoders, the authors in \cite{ref:DLsys} and \cite{ref:DLE2Esync} separately implemented the timing synchronization and the sampling time synchronization.
With DNN approaches, the authors in \cite{ref:DL-PDCFO} investigated the packet detection.
In \cite{ref:ELM-FTS}, an extreme learning machine (ELM) was employed to estimate the residual symbol timing offset (STO). From \cite{ref:CSIlearn,ref:DLsys,ref:DLE2Esync,ref:DL-PDCFO,ref:ELM-FTS}, previous studies suggested that the performance improvements for synchronization can be achieved by employing supervised learning.

Although supervised learning-based TS could provide performance improvements, the implementation bottleneck, essentially, will be encountered in the training data collection. For classification tasks such as TS, training data set is consist of the collected samples and their corresponding learning labels (e.g., labels by STOs). In the context of realistic wireless communication systems, collecting training data is time and labor consuming, especially for the supervised learning-based TS. For the supervised learning-based TS learner, the implementation bottleneck of training data collection and its correspondingly possible solution mainly exist in the following aspects:
\begin{enumerate}
  \item
        Precisely {labeling} the ground truth {labels} of TS is impractical to the receiver in realistic wireless communication systems. For one thing, due to unknown attributes of wireless channels, e.g., STOs, CFOs, and Channel Impulse Response (CIRs), the availability and optimal timing index of the received samples cannot be judged correctly. For another, conventional digital receivers distinguish the start point of collected samples according to the sampling grid. Namely, the received samples on off-grid are bound to be lost, making it challenging to precisely label the collected samples. As a result, the correlation between the estimated-based labels and the ground truth labels has been weakened, leading to the degradation of learning efficiency for the supervised learning-based TS learner.
        \emph{To solve this issue, desired training samples and their corresponding learning labels can be generated by computers, tackling the challenge of generating the ground truth labels in TS.
        Superior to the collected data, the computer-aided generated training data could perfectly know the generation parameters and methods (e.g., STO, CFO, path delays, etc.) while the former one cannot.}
  \item
        The completeness of training data collection is hindered in realistic wireless communication systems. In the long term, the realistic wireless channel environments are mainly affected by the climate or weather factors, such as the alternation of spring, summer, autumn, and winter, the temperature variation in the morning and evening, and the environments variation in rainy and sunny days. In the short term, the realistic wireless channel environments are mainly affected by obstacles in the cellular, such as vehicles and pedestrians. Thus, the maximum multi-path delay is time-varying, which makes it impractical to collect complete training data (including training samples and their corresponding learning labels) from the realistic wireless channels.
        \emph{To alleviate this challenge, more complete training samples can be generated by computers\cite{ref:BBD}, alleviating the completeness problem of the data collection in a realistic wireless communication. The insufficiency of training data may reduce the statistical efficiency of the supervised learning-based TS learner, while computers can generate rich data that inclose the changes of TS with environments. Namely, by employing a large number of wireless communication scenarios, the computer-aided approach can generate data more sufficiently than collect the desired data from realistic wireless communication systems. Especially, the generated data features are more conducive and faster to be obtained than those cannot be directly obtained by digital receivers~\cite{ref:SDIoT}, avoiding the time and labor consumptions.}
  \item
        Storing and transmitting the collected large-scale training data may cause substantial disasters. On the one hand, the supervised learning-based TS learner requires sufficient data to improve the adaptability to the changing wireless environments.
        However, large-scale data collection inevitably leads to great difficulty in storage capacity.
        On the other hand, large-scale training data transmission will bring unacceptable bandwidth requirements to the current wireless communication systems. Given the rapidly increasing demand for bandwidth resources, limited transmission bandwidth is mainly taken by the business data of users. Namely, it is unlikely to set aside valuable bandwidth resources merely for sending massive collected samples for network learning.
        \emph{To mitigate the pressures of storing and transmitting the collected large-scale training data for the TS, a computer-aided approach can deploy the corresponding algorithm of data generation at the local devices. By providing the constraints of data features, the desired data can be generated rapidly and released temporarily, {which alleviates the pressure from the requirement of huge storage space}. {Besides,} it is not necessary to transmit data for {network training, avoiding the large bandwidth requirement of} data transmission for supervised learning-based TS learners.}
\end{enumerate}

In addition, the current supervised learning-based TS learners relying on the DNN models (e.g., \cite{ref:DLsys,ref:DLE2Esync,ref:DL-PDCFO}) have achieved remarkable success in improving the TS performance. However, several public problems are still required to be solved, such as complex parameter tuning, long-time training, and local optimal solution \cite{ref:DL-CSIfeedback}. Unlike the DNN-based approaches, ELM is one kind of the single hidden layer feed-forward neural networks (SLFNs) without the requirement of gradient back-propagation \cite{ref:ELM1}, which may result in less time-consuming {for} training the TS learner.

{Inspired by the advantages of computer-aided approaches and ELM networks, we propose the computer-aided training strategy for ELM-based TS in this paper}, which aims to tackle the bottleneck facing in the training data collection and enhance the generalization performance of the designed TS learner.
The main contributions of our work are summarized as follows:
\begin{enumerate}
\item
    We develop a computer-aided approach to generate the training data for the ELM-based TS network in OFDM systems. {As a study case of this paper}, by exploring the partial properties of CIRs, some important data features of realistic wireless channels can be rapidly captured, e.g., the root-mean-square (r.m.s.) delay. Compared with classic TS schemes for training data collection, the proposed scheme has the ability to reduce the consumption of extracting the available and desired training data from the raw received data, i.e., less time- and labor-consuming. Meanwhile, due to the good controllability of computer-aided approaches, as many realistic situations as possible can be simulated with the assistance of computers, such that increases the completeness of training data sets. In addition, by generating training data at the local device, the proposed scheme reduces the burden of storage space and transmission bandwidth in the training data collection.
\item
    We exploit the rough features of CIRs (i.e., the expected maximum multi-path delay) to design the training strategy {for the ELM-based TS network}.
    Particularly, the loose constraint- and flexible constraint-based training strategies are derived from exploring the statistical information of multi-path delays, which aims to increase the generalization performance of TS learners and the correlation between the generated learning labels with the ground truth labels.
    The proposed training strategies inherit the advantage of the training strategy of \cite{ref:ELM-labelTS} (i.e., increasing the possibility of locating STOs into the ISI-free region) and also well adapt themselves to the volatile ground truth labels (i.e., offering a  strong correlation between the generated learning labels and the ground truth labels).
    With this, the designed training strategy can improve the generalization performance of the designed TS learner compared with that of \cite{ref:ELM-labelTS}. Moreover, sufficient training data can be rapidly produced with the assistance of the computer-aided approach, which avoids collecting the training data from realistic systems.

\item
    We form a {lightweight} TS framework for the OFDM systems by fusing the classic synchronizer and the conventional ELM, i.e., the ELM-based TS network.
    Particularly, the classic TS scheme is performed to directly extract the relevant features of timing information from the raw received samples.
    By this means, ELM can easily learn the transformation between the input features and the learning labels, even without the deeper hidden layers being used for fusing the implicit data features.

\end{enumerate}
\begin{table}[t]
\caption{Notations}\label{Note:sym}
\centering
\renewcommand\arraystretch{1.25}
\begin{tabular}{|c||c|}
\hline
\textbf{Notation} & \textbf{Description} \\ \hline
$\mathbb{R}^{M\times N}$ & $M$-by-$N$ dimensional real matrix space \\ \hline
$\mathbb{C}^{M\times N}$ & $M$-by-$N$ dimensional complex matrix space \\ \hline
$\mathbb{E}\{\cdot\}$ & Expectation operation\\ \hline
$e$ & Constant $e=2.71828\ldots$ \\ \hline
$\mathbf{0}_{M\times N}$ & $M$-by-$N$ dimensional all-zeros matrix \\ \hline
$\mathbf{1}_{M\times N}$ & $M$-by-$N$ dimensional all-ones matrix \\ \hline
$\mathds{I}_{N}$ & $N$-by-$N$ dimensional Identity matrix \\ \hline
$[\mathbf{X}]_{a,b}$ & Entry $(a,b)$ of matrix $\mathbf{X}$ \\ \hline
$\mathbf{X}_{a:b}$ & Sub-vector of $\mathbf{X}$ with entries $a$ to $b$ \\ \hline
$\mathcal{{CN}}\left(\mu,{\sigma}^2\right)$ & \makecell[c]{Circularly symmetric complex Gaussian\\(CSCG) distribution}\\ \hline
$\mathcal{U}(a,b)$ & Uniform distribution \\ \hline
$\left(\cdot\right)^{\ast}$ &  Conjugate operation\\ \hline
$\left(\cdot\right)^{\textit{T}}$ & Transpose operation\\ \hline
$\left(\cdot\right)^{\textit{H}}$ & Hermitian transpose \\ \hline
$\left(\cdot\right)^{\dag}$ & Moore--Penrose generalized inverse operation\\ \hline
$|\cdot|$ & Absolute operation \\ \hline
${\left\|\cdot\right\|_2}$ & $\ell_2$-norm operation\\ \hline
$\lceil\cdot \rceil$ & Integer ceiling operation\\ \hline
$\log_{10}(\cdot)$ & Logarithm base 10 \\ \hline
$\Pr\left(\cdot\right)$ & Probability of an event \\ \hline
\end{tabular}
\end{table}
The rest of this paper is organized as follows. For notation clarity, the meaning of symbols that have been adopted in this paper is listed in Table~\ref{Note:sym}. Section \ref{S:II} briefly describes the system model and problem formulation. Section III provides the computer-aided training strategies for the ELM-based TS network and the implementation of the proposed ELM-based TS network. Section \ref{S:IV} comprehensively evaluates the TS performance of the proposed computer-aided training strategy for ELM-based TS network, and is followed by conclusions in Section \ref{S:V}.
\section{System Model}\label{S:II}
\begin{figure}[t]
\centering
\includegraphics[width=0.5\textwidth]{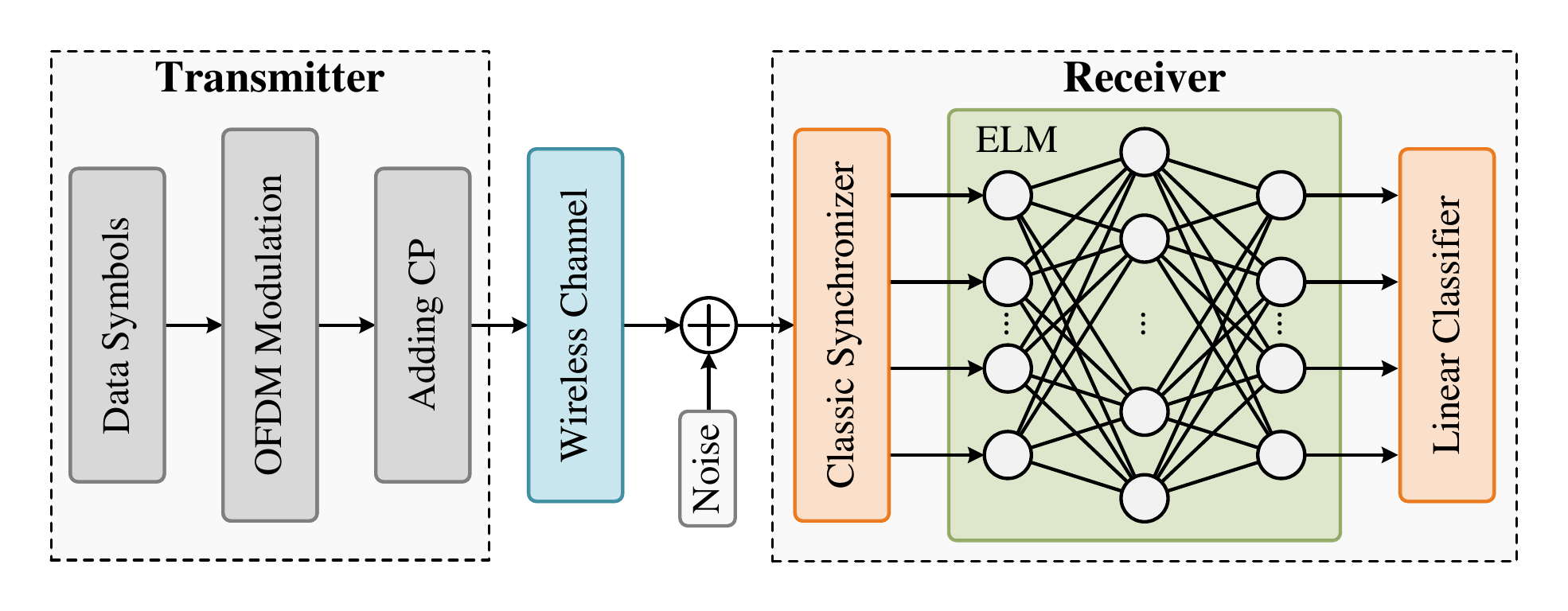}\\
\caption{System model.}\label{Fig:sys}
\end{figure}
We consider an OFDM system with $N$ sub-carriers, which is plotted in Fig~\ref{Fig:sys}. According to the inverse fast Fourier transform (IFFT), the frequency-domain symbol $\mathbf{d}\in \mathbb{C}^{N\times1}$ is first transformed into the time domain, and the OFDM symbol $\mathbf{s}\in \mathbb{C}^{N\times1}$ in the time domain is expressed as
\begin{equation}\label{EQ:1}
{\bf{s}} = {{{\bf{F}}_N}{\bf{d}}},
\end{equation}
where {$[\mathbf{s}]_{k,1} = s_k,k\in\{0,1,\cdots,N-1\}$} and $\mathbb{E}\{|s_k|^2\}=\sigma^2_d$ with $\sigma^2_d$ being {data power}. ${\bf{F}}_N\in\mathbb{C}^{N\times N}$ stands for the $N$-by-$N$ dimensional IDFT matrix and $[{\bf{F}}_N]_{k,l}=\frac{1}{{\sqrt N }}{e^{j \frac{{2\pi kl}}{N}}}$, ${k,l}\in\{0,1,\cdots,N-1\}$.
{The} cyclic prefix (CP) addition matrix $\mathbf{G}_{\mathrm{cp}}\in \mathbb{R}^{N+L_c\times N}${ is employed to denote the operation of adding} the CP for each OFDM symbol \cite{ref:CPMat}, where $L_c$ represents the length of CP. Then, the transmitted signal vector is given by
\begin{equation}
\mathbf{x} = \mathbf{G}_{\mathrm{cp}}\mathbf{s},
\end{equation}
where $\mathbf{x}$ is the $N_u \times 1$ transmitted signal vector with $N_u = N+L_c $, and
\begin{equation}\label{EQ:CPmat}
\mathbf{G}_{\mathrm{cp}} = \left[ {\begin{array}{*{20}{c}}
{{\mathbf{0}_{{L_c} \times {N - {L_c}}}}}&{{\mathds{I}_{{L_c}}}}\\\\
{{\mathds{I}_N}}&{{\mathbf{0}_{N \times {L_c}}}}
\end{array}} \right].
\end{equation}
By denoting $\mathbf{x} = [x_0, x_1,\cdots, x_{N_u}]^T$, the received signal $r_n$ over the multi-path fading channel is described as
\begin{equation}\label{EQ:rxdata}
r_n =\sum\limits_{l = 1}^L {h_lx_{n - \tau-\tau_{l} } {e^{j\frac{2\pi}{N} \varepsilon { n } }} }  + w_n, {n=0,1,\cdots,N_{u-1}},
\end{equation}
where $L$, $h_l$, $\tau$, $\tau_l$, and $\varepsilon$ stand for the taps of propagation channel, the channel coefficient of $l$th path, the unknown STO, the path delay of $l$th path, and the CFO being normalized by sub-carrier space, respectively. In \eqref{EQ:rxdata}, $w_n$ is the additive white noise satisfying the independent {Circularly symmetric complex Gaussian (CSCG)} distribution with zero-mean and variance $\sigma^2$. {Without loss of generality}, all multi-path delays are assumed to be shorter than the length of CP, i.e., $\tau_{l+1}>\tau_{l}, l=1,2,\cdots,P$ \cite{ref:MPDS}.
In the TS stage, the STO estimation is conducted, denoted as $\widehat{\tau}$. Particularly, we adopt an ELM network to combine with the classic synchronizer to enhance the TS for the OFDM systems.

One goal of our work is to generate more complete and accurate learning labels for TS, compared with the traditional labeling of learning labels for the collected training data in realistic wireless communication scenarios. Another goal is to avoid seriously occupying the bandwidth resource and storage space caused by the transmission and storage of collected training data. The computer-aided approach can appoint the generation method of training data at local devices, without additionally transmitting the collected data over the wireless channel. In addition, the training data can be timely generated without storing them and discarded promptly. To this end, the occupations of bandwidth resource and extra storage space could be avoided, inspiring us to develop computer-aided solutions.

\section{Computer-Aided Training Strategy for ELM-Based TS Network}
In this section, we first characterize the existing strategy from the perspective of obtaining learning labels, then elaborate the computer-aided strategy and finally present the designed TS framework for the OFDM systems. For the sake of clarity, a $K$-length learning label, denoted as $\mathbf{T}\in\mathbb{C}^{K\times1}$, can be represented by using a time-indexed sequence, i.e.,
\begin{equation}
\mathbf{T} = [T_0,T_1,\cdots,T_j,\cdots,T_{K-1}]^T.
\end{equation}

\subsection{Insights of Existing Strategy}
Deploying ML has become a promising solution to enhance/replace the conventional signal processing \cite{ref:Bellfurture}, however, ML-based signal processing still faces significant challenges with determining the estimated labels for training samples, and the ELM-based TS network is no exception.
\subsubsection{{Case I}}
If the information of CIRs are perfectly known to the receiver, three types of existed learning-label designs are mentioned and compared in \cite{ref:ELM-labelTS}, which are listed as follows:
\begin{itemize}
\item \textit{One-hot Coding:} When the expected maximum timing offset is known to be less than the length of OFDM symbol (i.e., $N$), the correct start location of preamble is employed to design the learning label. By utilizing the one-hot coding, the label value is given by \cite{ref:ELM-FS}
    \begin{equation}\label{EQ:Lone}
    {T_j} = \left\{ \begin{array}{l}
    1,\ j = \tau  + {L_c}\\
    0,\ \textrm{others}
    \end{array} \right.,
    \end{equation}
    where $j\in\{0,1,\cdots,K-1\},K=N_u=N+L_c$, due to the existence of CP. The vector form of this learning label is $\mathbf{T}=[\bm{0}_{1\times\tau+L_c},1,\bm{0}_{1\times K-\tau-L_c-1}]^T$.
    Since the TS of OFDM system only requires that the STO estimation falls into the ISI-free region, the timing instants within the ISI-free region can be regarded as the ground truth labels.
    Without the use of cyclic suffix, the nonzero label feature points to the edge of the ISI-free region such that results in a weak correlation between the generated learning labels and truth ground labels.
\item \textit{Labeling ISI-free Midpoint:} With the use of known timing offsets and maximum multi-path delay, the midpoint of ISI-free region is exploited to design learning label \cite{ref:ELM-labelTS}. The midpoint of ISI-free region is denoted as $p_c=\lceil(\tau_P+L_c)/2\rceil$, which is used for designing the label value \cite{ref:ELM-labelTS}
        \begin{equation}\label{EQ:Ltwo}
        {T_j} = \left\{ \begin{array}{l}
        1,\ j = \tau  + p_c\\
        0,\ \textrm{others}
        \end{array} \right..
        \end{equation}
    and its vector form is $\mathbf{T}=[\bm{0}_{1\times\tau+p_c },{1},\bm{0}_{1\times K-\tau-p_c-1}]^T$.
      By allocating nonzero value corresponding to the midpoint of ISI-free region, this label increases the correlation between the nonzero label feature and the ground truth label, such that increases the possibility of residing the estimated STO to the ISI-free region.
\item \textit{{Labeling ISI-free Region:}} By completely exploring the prior information of ISI-free region, the label value is given by \cite{ref:ELM-labelTS}
\begin{equation}\label{EQ:Lthree}
{T_j} = \left\{ \begin{array}{l}
1,\ \tau  + {\tau _{P }} \le j \le \tau  + {L_c}\\
0,\ \textrm{others}
\end{array} \right.,
\end{equation}
and its vector form can be expressed as
$\mathbf{T}=[\bm{0}_{1\times\tau+\tau_P},\bm{1}_{1\times L_c-\tau_P+1},\bm{0}_{1\times K-\tau-L_c-1}]^T$. By setting nonzero label values corresponding to the whole ISI-free region, this label provides extra information to increase the correlation between the nonzero label features and the timing indexes of ISI-free region, which greatly enhances the ELM learning. As a result, the ELM-based TS network combined with the label in \eqref{EQ:Lthree} has a lower error probability of TS than that combined with the label in \eqref{EQ:Ltwo}.
\end{itemize}

However, the complete real-time multi-path delays are difficult to be pre-known at the receiver, so the completeness of learning labels in \eqref{EQ:Ltwo} and \eqref{EQ:Lthree} are facing challenge with the varied maximum multi-path delay.
%In realistic wireless communication systems, the main issue of the learning labels  is that: the received samples are unknown to the receiver, which makes it challenging to generate the correct learning labels.
\subsubsection{Case II}
If the information of CIRs cannot be perfectly known at the receiver, employing the classic TS schemes to collect the training data is a simple and natural option.
By taking the TS scheme of \cite{ref:SC} as an example, the learning labels can be obtained according to the timing metric.
With the received samples of $r_n$, the timing metric $M_j$, at $j$th lag of correlation, is given by
\begin{equation}\label{EQ:etm}
{M_j} = \frac{{{{\left| {\sum\limits_{d = 0}^{\frac{N}{2}} {r_{j + d}^ * {r_{j + \frac{N}{2} + d}}} } \right|}^2}}}{{{{\left| {\sum\limits_{d = 0}^{\frac{N}{2}} {r_{j + \frac{N}{2} + d}^ * {r_{j + \frac{N}{2} + d}}} } \right|}^2}}},
\end{equation}
and then $\widehat{\tau}$ is obtained by $\widehat{\tau}  = \mathop {\arg \max }\limits_{0 \le j \le K - 1} \left\{ {{M_j}} \right\}$.
With the use of $\widehat{\tau}$, the generated learning label can be expressed as
\begin{equation}\label{EQ:CCL}
{T_j} = \left\{ \begin{array}{l}
1,\ j = \widehat{\tau} \\
0,\ \textrm{others}
\end{array} \right..
\end{equation}
This label is {direct}, and its vector form is denoted as $\mathbf{T}_{\textrm{CT}}\in\mathbb{R}^{K\times1}$ and given by $\mathbf{T}_{\textrm{CT}}=[\bm{0}_{1\times\widehat{\tau}},{1},\bm{0}_{1\times K-\widehat{\tau}-1}]^T$.
We assume that, $N_t$-samples of $\mathbf{r}_m\in\mathbb{C}^{K+N\times1}$ are required for the model training. In addition, $\widehat{\tau}_m$ and ${\mathcal{R}^{(m)}_{\textrm{ISI-free}}}$  are denoted as the estimated STO and the ISI-free region for $\mathbf{r}_m$, wherein ${\mathcal{R}^{(m)}_{\textrm{ISI-free}}}$ is given by ${\mathcal{R}^{(m)}_{\textrm{ISI-free}}}=\tau^{(m)}+\{\tau^{(m)}_P,\tau^{(m)}_P+1,\cdots,L_c\}$. Then, the metric for the accuracy of estimated learning labels can be expressed as
\begin{equation}\label{EQ:Pr1}
{P_{\textrm{label}}} = \frac{1}{{{N_t}}}\sum\limits_{m = 1}^{{N_t}} {\theta \left( {{{ \widehat{\tau} }_m}} \right)},
\end{equation}
where
\begin{equation}\label{EQ:Pr2}
\theta \left( {{{ \widehat{\tau} }_m}} \right) = \left\{ \begin{array}{l}
1,\ {{ \widehat{\tau} }_m} \in {\mathcal{R}^{(m)}_{\textrm{ISI-free}}}\\
0,\ {{ \widehat{\tau} }_m} \notin {\mathcal{R}^{(m)}_{\textrm{ISI-free}}}
\end{array} \right..
\end{equation}
The above $\theta \left( {{{ \widehat{\tau} }_m}} \right)$ indicates whether the estimated STO (i.e., $\widehat{\tau}_m$ for generating learning label) is correct or not.
\begin{remark}
According to \eqref{EQ:Pr2}, the obtained labels are partially flawed due to the TS errors. That is, when the estimated $\widehat{\tau}$ points to the edge or outside of the ISI-free region, the correlation between the estimated labels (i.e., $\widehat{\tau}$) and the timing indexes of ISI-free region is seriously impaired. The same issues can be found in other classic TS schemes. To guarantee the efficiency of training data, a large-scale of training samples are required to be collected for selection.
For example, if the accuracy of estimated learning labels is $P_{\textrm{label}}$, at least $N_t/P_{\textrm{label}}$ training samples are required to obtain $N_t$ of effective learning labels.
As a result, by leveraging the classic TS scheme, the training data collection is unfriendly to limited storage space and transmission bandwidth resources.
\end{remark}
%{\textbf{Remark 2:}} It is worth noting that, correctly collecting the training data is not only a matter of time.
%To obtain complete data sets, the crucially steep challenges facing in the training data collection are described as follows.
%\begin{enumerate}
%  \item Owing to the limited resolutions in the time domain, frequency domain, etc, it is difficult for real digital receivers to obtain the off-grid data. This inevitably results in some important data loss in data collection.
%  \item In realistic wireless communication systems, the profile of CIRs is time-varying, so the training data collection is difficult and time-consuming. Meanwhile, it is impractical to obtain all CIR profiles. This greatly reduces the completeness of training data.
%  \item During the training data collection, large-scale data samples result in the curse of storage space. Meanwhile, massive data samples are employed to train a neural network, bringing the pressure of bandwidth resources if the transmission of large-scale data is needed.
%\end{enumerate}
To alleviate the above challenge, we develop computer-aided training strategies, which is presented in Section III-B in detailed.

\subsection{Computer-Aided Strategy}
It could be obtained from \eqref{EQ:Pr2} that, the accuracy of estimated learning labels is heavily influenced by the TS errors. This is because the performance of classic TS schemes is vulnerable to the impacts of CIRs and noise. To avoid incorrectly labeling the training samples, computers can be utilized to generate the training data (including the training samples and their corresponding learning labels) at the local device, since the computer-aided generated training data is controllable and known to the transceiver.
In the following, we mainly discuss the strategy of improving the completeness with respect to the varied maximum multi-path delays.

\subsubsection{\textbf{Loose Constraint}}
The completeness of training data is mainly impaired by the profile of CIRs. In the TS stage, the key insight is to avoid the ISI, i.e., locating the timing indexes in the ISI-free region. To this end, we propose a loose constraint for the profile of CIRs to improve the completeness of training data. In other words, the loose constraint is utilized to generate the training samples and design the learning labels. The key idea of the proposed loss constraint is motivated by these two factors: 1) there is a time range for the realistic maximum multi-path delay to dynamically change; 2) statistical information of multi-path delays in a special geographical environment is easier to be obtained compared with acquiring the realistic multi-path delays themselves. The main \:motivation is that, the r.m.s. delay is usually measured prior to the establishment of terrestrial mobile communication systems e.g., the wireless communication standards for the 4G and 5G communication systems~\cite{ref:3GPP4G5G,ref:3GPP5G}, and so that the r.m.s. delay can constrain the maximum multi-path delay to a specific time range with high probability.

Here, we denote $\mathcal{L}$ as the loose constraint for the maximum multi-path delay in the training stage, which is utilized to generate the maximum multi-path delays via computer-aided data generation. Specifically, to avoid ISI, $\tau_P<\mathcal{L}<L_c$ is given, and $\mathcal{L}$ is sufficient to tolerate the increase of $\tau_P$. Although the underlying mechanism of $\mathcal{L}$ is similar to the CP in the aspect of avoiding ISI, this loose constraint is only used to generate training data in the training stage, which does not need extra overhead to change the frame structure. For $N_t$ samples of $\{\mathbf{r}_m\}^{N_t}_{m=1}$ being required, ${\mathcal{R}^{(m)}_{\textrm{ISI-free}}}=\tau^{(m)}+\{\mathcal{L},\mathcal{L}+1,\cdots,L_c\}$ is denoted as the target ISI-free region (i.e., target STOs for designing learning label) for each sample of $\mathbf{r}_m$. Correspondingly, the ground truth label for the realistic communication systems are denoted as $\mathcal{R}_{\textrm{truth}}=\tau+\{\tau_P,\tau_P+1,\cdots,L_c\}$. Then, for the case where $\tau^{(m)}=\tau$, the correlation between ${\mathcal{R}^{(m)}_{\textrm{ISI-free}}}$ and $\mathcal{R}_{\textrm{truth}}$ is described as
\begin{equation}\label{EQ:LC_Corr}
\mathcal{R}^{(m)}_{\textrm{ISI-free}} \subset \mathcal{R}_{\textrm{truth}},\ \forall m\in\{1,2,\cdots,N_t\}.
\end{equation}
According to \eqref{EQ:LC_Corr}, since $\mathcal{L}$ is sufficient to the possible maximum multi-path delays, $\mathcal{R}^{(m)}_{\textrm{ISI-free}}$ always contains the correct learning labels that correspond to the ground truth labels. This improves the completeness of training data.
For example, when $L_c=32$ and $\tau_{P}$ possibly ranges from 16 to 23, $\mathcal{L}$ can be set as $26$ for generating each training sample.

The works in \cite{ref:ELM-labelTS} verified that the ELM network with the learning label in \eqref{EQ:Ltwo} has good generalization against the multi-path delays, and so we mainly focus on the improvement for the learning label in \eqref{EQ:Lthree}. With the use of $\mathcal{L}$, the improved learning label is written as
\begin{equation}\label{EQ:Lv3new}
{T_j} = \left\{ \begin{array}{l}
1,\ \tau+{\cal L} \le j \le \tau  + L_c\\
0,\ \textrm{others}
\end{array} \right.,
\end{equation}
and its vector form is denoted as $\mathbf{T}_{\textrm{LC}}\in\mathbb{R}^{K\times1}$ and given by $\mathbf{T}_{\textrm{LC}}=[\bm{0}_{1\times\tau+{\cal L}},\bm{1}_{1\times L_c-{\cal L}+1},\bm{0}_{1\times K-\tau-L_c-1}]^T$.
Since $\tau_{P}$ is smaller than $\cal L$, the supervised learning-based TS learner combined with $\mathbf{T}_{\textrm{LC}}$ can actively avoid the potentially discriminatory behaviors that degrade the generalization performance.

However, according to \eqref{EQ:LC_Corr}, the loose constraint-based strategy for training data generation lacks a sufficient number of training labels that have a strong correlation with the ground truth labels, since $\mathcal{R}_{\textrm{truth}}\neq\mathcal{R}^{(m)}_{\textrm{ISI-free}}$ or $\mathcal{R}_{\textrm{truth}}\not\subset\mathcal{R}^{(m)}_{\textrm{ISI-free}}$. On the one hand, if $\mathcal{R}_{\textrm{truth}}\equiv\mathcal{R}^{(m)}_{\textrm{ISI-free}}$, the training labels will show a strong correlation with the ground truth labels, such that facilitates the performance of TS learners. On the other hand, if $\mathcal{R}_{\textrm{truth}}\subset\mathcal{R}^{(m)}_{\textrm{ISI-free}}$, the training data set can be regarded as the complete set of possible maximum multi-path delays, such that improves the statistical efficiency of TS learner.
As a result, the quality and efficiency of the supervised learning-based TS leaner are degraded.
\subsubsection{\textbf{Flexible Constraint}}
Flexible constraint has the ability to improve the adaptability to the various demands in practical applications, such as the flexible constraint length Viterbi decoder \cite{ref:FConst}. For the supervised learning-based TS leaner, the key insight of introducing the flexible constraint into the design of training strategy is that: increasing the correlation between the training labels and the ground truth labels is helpful to improve the statistical power of the supervised learning-based TS leaner. To this end, we propose the flexible constraint for generating the training data on the basis of loose constraints.

By denoting $\widetilde{\cal L}$ as the flexible constraint for the maximum multi-path delay in the training stage, the target STOs for designing learning label is denoted by ${\mathcal{R}^{(m)}_{\textrm{ISI-free}}}=\tau^{(m)}+\{\widetilde{\mathcal{L}}^{(m)},\widetilde{\mathcal{L}}^{(m)}+1,\cdots,L_c\}$,
with $\widetilde{\mathcal{L}}^{(m)}$ being distribution as $\widetilde{\mathcal{L}}^{(m)}\sim\mathcal{U}(L_c/2,\mathcal{L})$. Similarly, we denote $\mathcal{R}_{\textrm{truth}}=\tau+\{\tau_P,\tau_P+1,\cdots,L_c\}$ as the ground truth labels for the realistic communication systems. Then, for the cases where $\tau_P>L_c/2$ and $\tau^{(m)}=\tau$, the correlation between ${\mathcal{R}^{(m)}_{\textrm{ISI-free}}}$ and $\mathcal{R}_{\textrm{truth}}$ is expressed as
\begin{equation}\label{EQ:FC_Corr}
\mathcal{R}^{(m)}_{\textrm{ISI-free}} \supseteq \mathcal{R}_{\textrm{truth}},\ \forall m\in\{1,2,\cdots,N_t\}.
\end{equation}
With this, if the amount of training samples is sufficient, the given $\mathcal{R}^{(m)}_{\textrm{ISI-free}}$ can be approximately regarded as the complete set of possible maximum multi-path delays in some cases. As for the cases where $\tau_P<L_c/2$ and $\tau^{(m)}=\tau$, the correlation given in \eqref{EQ:FC_Corr} can be rewritten as $\mathcal{R}^{(m)}_{\textrm{ISI-free}} =\mathcal{R}_{\textrm{truth}},\ \exists m\in\{1,2,\cdots,N_t\}.$ Wherein $\tau_P<L_c/2$ can be regraded as a case of $\mathcal{L}$, i.e., $\mathcal{L}=L_c/2$. Therefore, the proposed flexible constraint inherits the advantages of the proposed loose constraint and also improves the correlation between the training labels and ground truth labels.

By leveraging $\widetilde{\mathcal{L}}$, the improved learning label is given by
\begin{equation}\label{EQ:Lv3new}
{T_j} = \left\{ \begin{array}{l}
1,\ \tau+\widetilde{\cal L} \le j \le \tau  + L_c\\
0,\ \textrm{others}
\end{array} \right..
\end{equation}
and its vector form is denoted as $\mathbf{T}_{\textrm{FC}}\in\mathbb{R}^{K\times1}$ and given by  $\mathbf{T}_{\textrm{FC}}=[\bm{0}_{1\times\tau+{\cal L}},\bm{1}_{1\times L_c-{\cal L}+1},\bm{0}_{1\times K-\tau-L_c-1}]^T$.
Due to the fact that $\widetilde{\mathcal{L}}\sim\mathcal{U}\left(L_c/2, \mathcal{L}\right)$, the supervised learning-based TS learner can learn more training data that may appear in practical environments. This improves the correlation between the training labels and the ground truth labels, and so that improves the statistical power of TS learners. Moreover, instead of collecting sufficient data features from realistic systems, a large number of rich and diverse data features can be rapidly generated by $\widetilde{\mathcal{L}}$ at local devices. Thus, not only the occupations of storage space and transmission bandwidth caused by frequently collecting desired data features are correspondingly avoided, but also more memory traces to the neural network can be retained against the changing profile of CIRs.
\begin{remark}
It is impractical to completely collect training data from realistic systems since the received samples are unavailable before the completion of synchronization. As for the supervised learning-based TS learner, it is available for local devices (i.e., employing the Central Processing Unit) to generate efficient and more complete training data. By this means, once the computer-aided approach is adopted, the training samples and their corresponding learning labels can be timely generated and released. Compared with the training data collected by the classic TS schemes, more complete and accurate training data can be generated at the local device, and thus greatly reduces the storage consumption and bandwidth occupation of collecting large-scale data from the realistic systems. It is worth mentioning that, the more complete training data features we obtain, the more statistical efficiency the neural network will own. Thus, a good generalization capability of ELM-based TS network and low occupations of storage space and transmission bandwidth can be achieved by the proposed strategies of loose constraint and flexible constraint.
\end{remark}
\subsection{ELM-Based TS Network Architecture}\label{S:III}

The proposed ELM-based TS network consists of two phases, i.e., a pre-processing phase of feature extraction and an ELM batch learning phase. In the pre-processing phase, the received signals are processed via a classic TS scheme for feature extraction. The timing metrics calculated from the raw received samples are regarded as extracted features, which are then filled up for the ELM batch learning phase. Within timing metrics, the extracted features should be normalized by the Frobenius norm (i.e., $\ell_2$-norm). Following the pre-processing phase, the ELM batch learning phase is carried out to learn the mapping relationship between the input features and the designed learning labels as desired. To achieve our desired, the training data sets are generated with the proposed computer-aided training strategy (i.e., the loose constraint- and flexible constraint-based strategies in Section III-B), which are controllable, distinct, and reliable for the ELM learning.

For the convenience of expression, preliminary preparation of ELM-based TS network is characterized below. Selecting $N_h$-numbers of hidden layer nodes, we denote $\bm{\mathcal{D}}_{\textrm{ca}}=\{(\mathbf{r}_m,\mathbf{T}_m)|\mathbf{r}_m\in{\mathbb{C}}^{N_w\times1}, \mathbf{T}_m\in{\mathbb{R}}^{K\times1} , m=1,2,\cdots,N_t\}$ as the original training data set. Wherein $N_w$ denotes the size of the observation window for received signals, and $K$ is the size of the observation window for timing metrics. Specifically, the received samples of  $\{\mathbf{r}\}^{N_t}_{m=1}$ are obtained according to \eqref{EQ:1}--\eqref{EQ:rxdata} without interferences of CFO and noise. The generation of learning labels (i.e., $\{\mathbf{T}\}^{N_t}_{m=1}$) is given in Section III-B. Besides, the architecture of the adopted ELM network is presented in Table \ref{Tab:S}.

\subsubsection{Offline Training}
\begin{table}[t]
\caption{Network Architecture}
\renewcommand\arraystretch{1.15}
\centering
\begin{tabu}{c|c|c|c}
\tabucline[1pt]{-}
%\diagbox{Setting}{Layer}
 & Input Layer & Hidden Layer & Output Layer \\ \tabucline[1pt]{-}
Neuronal Nodes      & \(K\) & \({N_h}\)    & \(K\)           \\ \hline
Activation Function & None       & Tanh        & None \\\hline
Normalization & $\ell_2$-norm       & None        & None \\
\tabucline[1pt]{-}
\end{tabu}
\label{Tab:S}
\end{table}
Algorithm 1 summarizes the process of offline training for the ELM-based TS network, and the detailed description is elaborated as follows.

\textbf{Step 1) Initial Feature Extraction:} Initially extracting the features using received samples $\{\mathbf{r}_m\}^{N_t}_{m=1}$ from the given training data set $\bm{\mathcal{D}}_\textrm{ca}$. By taking the classic synchronizer (i.e., TS scheme in \cite{ref:SC}) as an example, the specifical preamble consists of two identical parts. In addition, the observed sub-vector of $\mathbf{r}_m$ at $j$th lag of correlation, denoted as $\mathbf{r}_{m,j}=[\mathbf{r}_m]_{j:j+\frac{N}{2}}$, is expressed as
\begin{equation}\label{EQ:ObSlotdata}
\mathbf{r}_{m,j} ={\left[ {{r_{m,j}}, \cdots ,{r_{m,j + d}}, \cdots ,{r_{m,j + N - 1}}} \right]^T},
\end{equation}
where $j\le N_w-N+1$ and $N_w= 2N_u$ with the aim of observing a complete preamble at the receiver. Then, the timing metric is employed as the ELM input and given by
\begin{equation}\label{EQ:TMM}
{M_{m,j}} = \frac{{{{\left| {{P_{m,j}}} \right|}^2}}}{{{\frac{1}{4  }{\left| {{R_{m,j}}} \right|}^2}}},
\end{equation}
where $P_{m,j}$ and $R_{m,j}$ stand for the autocorrelation factor \cite{ref:SC} and the normalization factor \cite{ref:RR}, respectively, which are defined as
\begin{equation}\label{EQ:PandQ}
\left\{ \begin{array}{l}
{P_{m,j}} = \sum\limits_{d = 0}^{\frac{N}{2} - 1} {r_{m,j + d}^ * {r_{m,j + d + \frac{N}{2}}}} \\
{R_{m,j}} = \sum\limits_{d = 0}^{N - 1} {r_{m,j + d}^ * {r^{}_{m,j + d}}}
\end{array} \right..
\end{equation}
The iteration formula for $P_{m,j+1}$ is derived in \cite{ref:SC} to reduce the time complexity, i.e.,
\begin{equation}\label{EQ:itP}
{M_{m,j + 1}} = \frac{{{{\left| {{P_{m,j}} + r_{m,j + \frac{N}{2}}^ * {r^{}_{m,j + N}} - r_{m,j}^ * {r_{m,j + \frac{N}{2}}}} \right|}^2}}}{{{\frac{1}{4}{\left| {{R_{m,j}} + r_{m,j + N}^ * {r^{}_{m,j + N}} - r_{m,j}^ * {r^{}_{m,j}}} \right|}^2}}}.
\end{equation}
By buffering $K$-samples of $M_{m,j}$, the timing metric vector $\mathbf{M}_m\in\mathbb{R}^{K\times1}$ is constructed by
\begin{equation}\label{EQ:Mj}
\mathbf{M}_m=[M_{m,0},M_{m,1},\cdots,M_{m,K-1}]^T,\ K=N_u.
\end{equation}
To increase the accuracy of the classification and the learning efficiency, $\mathbf{M}_m$ is normalized by $\ell_2$-norm, which is expressed as
\begin{equation}\label{EQ:NFV}
\mathbf{\widetilde{g}}_m = \frac{\mathbf{M}_m}{{{{\left\| \mathbf{M}_m \right\|}_2}}},
\end{equation}
where $\mathbf{\widetilde{g}}_m\in\mathbb{R}^{K\times1}$ denotes the normalized timing metric vector. For the followed ELM batch learning, the above \eqref{EQ:1}--\eqref{EQ:rxdata} and \eqref{EQ:ObSlotdata}--\eqref{EQ:NFV} are repeated till $N_t$-samples of $\mathbf{\widetilde{g}}_m$ are achieved.

\textbf{Step 2) ELM Batch Learning:} Denoting $N_t$-samples of $\{(\widetilde{\mathbf{g}}_m,\mathbf{T}_m)|\widetilde{\mathbf{g}}_m\in{\mathbb{R}}^{K\times1}, \mathbf{T}_m\in{\mathbb{R}}^{K\times1} , m=1,2,\cdots,N_t\}$ as the training data for the ELM learning. To initialize the network, the input weight matrix $\mathbf{W}\in\mathbb{R}^{N_h\times K}$ and hidden bias vector $\mathbf{b}\in\mathbb{R}^{N_h\times1}$ are randomly selected from Gaussian variables, i.e., $\mathcal{N}(\mu,\sigma^2)$. In addition, the universal hyperbolic tangent (tanh) activation function is selected, since the classification task using tanh activation function has a lower classification error than other activation functions in general.

For each feed-in $\widetilde{\mathbf{g}}_m$, the hidden layer output matrix $\mathbf{H}_m$ is calculated according to
\begin{equation}\label{EQ:Hi}
{\mathbf{H}_m} = {f_\sigma }\left( {\mathbf{W} {\widetilde{\mathbf{g}}_m} + \mathbf{b}} \right),
\end{equation}
where $f_{\sigma}(\cdot)$ stands for the universal tanh activation function.
Repeating the above \eqref{EQ:Hi} till $N_t$-samples of $\mathbf{H}_m$ are collected. Next, on the basis of $N_t$-samples of $\left\{ {\left( {{\mathbf{H}_m},{\mathbf{T}_m}} \right)} \right\}_{m = 1}^{{N_t}}$ as attributes, the hidden layer output matrix $\mathbf{{H}}\in\mathbb{R}^{N_h\times N_t}$ and the label matrix $\mathbf{{T}}\in\mathbb{R}^{K\times N_t}$ are expressed as
\begin{equation}\label{Hmat}
\left\{ \begin{array}{l}
{\bf{ {H}}} = [{{\bf{H}}_1},{{\bf{H}}_2}, \cdots ,{{\bf{H}}_{{N_t}}}]\\
{\bf{ {{T}}}} = [{{\bf{T}}_1},{{\bf{T}}_2}, \cdots ,{{\bf{T}}_{{N_t}}}]
\end{array} \right..
\end{equation}
Then, the output weight matrix $\bm{\beta}\in\mathbb{R}^{K\times N_h}$ is given by
\begin{equation}\label{EQ:beta}
\bm{\beta}  = {{\mathbf{T}}} {{\mathbf{H}}^\dag }.
\end{equation}
The above trained parameters (i.e., $\mathbf{W}$, $\mathbf{b}$, and $\bm{\beta}$) are saved for online deployment.
\begin{remark}
The {initial feature extraction} is vitally important to improve the learning capability of ML models \cite{ref:OAR}. As for simply models like ELM network, the cleverly designed feature extractor could actively extract the pertinent information of STOs from raw data (e.g., coarse STO estimation), which benefits the ELM to {enhance learning effectiveness}. To this end, the classic TS scheme is employed to implement the feature extraction to capture timing metrics.
Also, since the normalization makes the features between different dimensions have a certain numerical comparison, to normalize the data in \eqref{EQ:NFV} before feeding it to the network is a good practice, which can greatly improve the accuracy of the classification.
\end{remark}
\begin{algorithm}[t]
\caption{Offline training} %算法的名字
\hspace*{0.02in} {\bf Input:} %算法的输入， \hspace*{0.02in}用来控制位置，同时利用 \\ 进行换行
\ \ Data set $(\mathbf{r}_m,\mathbf{T}_m),m=1,2,\cdots,N_t$. \\
\hspace*{0.02in} {\bf Output:} %算法的结果输出
Output weight matrix $\bm{\beta}$.
\begin{algorithmic}[1]
\State Set $N_h$-numbers of hidden layer nodes.
\State Randomly assign input weight matrix $\mathbf{W}$ and hidden bias vector $\mathbf{b}$.
\State Calculate timing metric ${\mathbf{M}}_m$ from $\{\mathbf{r}_m\}^{N_t}_{m=1}$ by \eqref{EQ:TMM}--\eqref{EQ:NFV}.
\State Compute hidden layer output $\mathbf{H}_m$ with the use of $\mathbf{M}_m$, $\mathbf{W}$, and $\mathbf{b}$ by \eqref{EQ:NFV}--\eqref{EQ:Hi}.
\State Set $\mathbf{H}=[\mathbf{H}_0,\mathbf{H}_1,\cdots,\mathbf{H}_{N_t}]$, $\widetilde{\mathbf{T}} =[\mathbf{T}_0,\mathbf{T}_1,\cdots,\mathbf{T}_{N_t}]$.
\State Estimate the output weight matrix $\bm{\beta}$ via $\bm{\beta} = \widetilde{\mathbf{T}}\mathbf{H}^{\dag}$.
\State \Return $\bm{\beta}$.
\end{algorithmic}
\end{algorithm}\label{T3}
\subsubsection{{Online Deployment}}
In this phase, a straightforward process of ELM-based TS network is carried out for online deployment. First, according to \eqref{EQ:1}--\eqref{EQ:rxdata} and \eqref{EQ:ObSlotdata}--\eqref{EQ:NFV}, the timing metric is calculated from received signals to capture the coarse timing feature, namely, $\widetilde{\mathbf{g}}\in\mathbb{R}^{N_w\times1}$. Next, the extracted $\widetilde{\mathbf{g}}$ is fed to the trained ELM network to capture the network output $\mathbf{O}\in\mathbb{R}^{K\times1}$, i.e.,
\begin{equation}\label{EQ:out}
\mathbf{O} = \bm{\beta}  {f_\sigma }\left( {\mathbf{W} \widetilde{\mathbf{g}} + \mathbf{b}} \right).
\end{equation}
By expressing $\mathbf{O}$ as $\{O_j\}^{K-1}_{j=0}$, the estimated STO is given by
\begin{equation}
\widehat{\tau}=\mathop {\arg \max }\limits_{0 \le j \le {K} - 1} \left\{ { {{O_j}} } \right\}.
\end{equation}

\section{Simulation Results}\label{S:IV}
In this section, numerical results are provided to evaluate the TS performance of the proposed computer-aided strategies (i.e., loose constraint- and flexible constraint-based strategies). Basic parameters and definitions involved in simulations are summarized in Section \ref{VA}. To indicate the effectiveness of the proposed strategies for generating training data, Section \ref{VB} provides the error probability of TS against the impacts of the accuracy and completeness of learning labels. Then, Section \ref{VC} presents the generalization analysis against the varying parameter of maximum multi-path delay, and a series of robustness analyses are provided in Section \ref{VD}. \emph{In the simulations, the testing channel models have not been utilized for the offline training phase, which aims to prove the feasibility of our study.}
\subsection{Parameter Setting}\label{VA}
In the simulations, we consider basic parameters as that $N=128$, $L_c=32$, $K=N_u=160$, $N_w=2N_u=320$, $N_h=8K$, and $N_t=10^{5}$.
The signal bandwidth for transmitting the collected data samples is considered as $\mathcal{W}=100$MHz \cite{ref:CTBW}. By employing the data type float32, a collected training sample is 4 Bytes. The transmitted preamble is generated according to the Zadoff-Chu sequence \cite{ref:TSZC} with the power of transmitted preamble being constant as $\sigma^2_p$, and $\sigma^2_d=\sigma^2_p$ is considered.
{A Rayleigh multi-path fading channel is considered, in which the exponential power delay profile is defined as} $\sigma^2_p = \frac{1}{\Sigma^{P}_{p=1}e^{-\eta\left(p-1\right)}}e^{-\eta\left(p-1\right)}$, with $\eta=0.2$.
Except for the testing maximum multi-path delay (i.e., $\tau_{P}$) that needs to discuss, the default parameter of $\tau_{P}$ is set as 20 sample delays. Correspondingly, the default loose constraint (i.e., $\cal L$) is set as 26 sample delays, and the flexible constraint (i.e., $\widetilde{\cal L}$) satisfies that $\widetilde{\mathcal{L}}\sim \mathcal{U}(16, 26)$.

Some definitions in the simulations are given as follows. The signal-to-noise-ratio (SNR) in decibel (dB)~\cite{ref:DL-CSIfeedback} is considered in the following simulations, i.e.,
\begin{equation}
\mathrm{SNR}_p = 10 \log_{10}\frac{\sigma^{2}_d}{\sigma^2_n}.
\end{equation}
The error TS means that the estimated STO points outside the ISI-free region, wherein the error probability of TS, denoted as $P_e$, is defined as
\begin{equation}\label{EQ:Pr}
P_e=\frac{Q_0}{Q_1},
\end{equation}
where $Q_0$ denotes the number of arriving preambles with error estimated STOs, and $Q_1$ denotes the total number of transmitted preambles.

For the sake of clarity, we denote ``DatCol'' as the existed strategy for collecting training data (including the training samples and the estimated labels), in which the collection of training data is conducted by the classic TS scheme (i.e., \cite{ref:CTS} for example), with $\mathrm{SNR}_p=20$dB.
And ``Ref \cite{ref:ELM-labelTS}'' is denoted as the strategy of \cite{ref:ELM-labelTS} for generating learning labels.
Correspondingly, ``Prop\_LC'' and ``Prop\_FC'' are denoted as the proposed loose constraint-based strategy and the proposed flexible constraint-based strategy for generating training data, respectively.
In addition, we adopt the TS scheme of \cite{ref:CTS}, denoted as ``Corr'', as the baseline in the simulations.
For a fair comparison, except for the comparative parameters, the given TS methods in each figure(e.g., Fig. 2 to Fig. 7) separately have the same parameter settings.

\begin{figure}[!ht]
\centering
    \subfigure[Impact of the accuracy of learning labels.]{
    \label{fig_comp1}
    \includegraphics[width=0.5\textwidth]{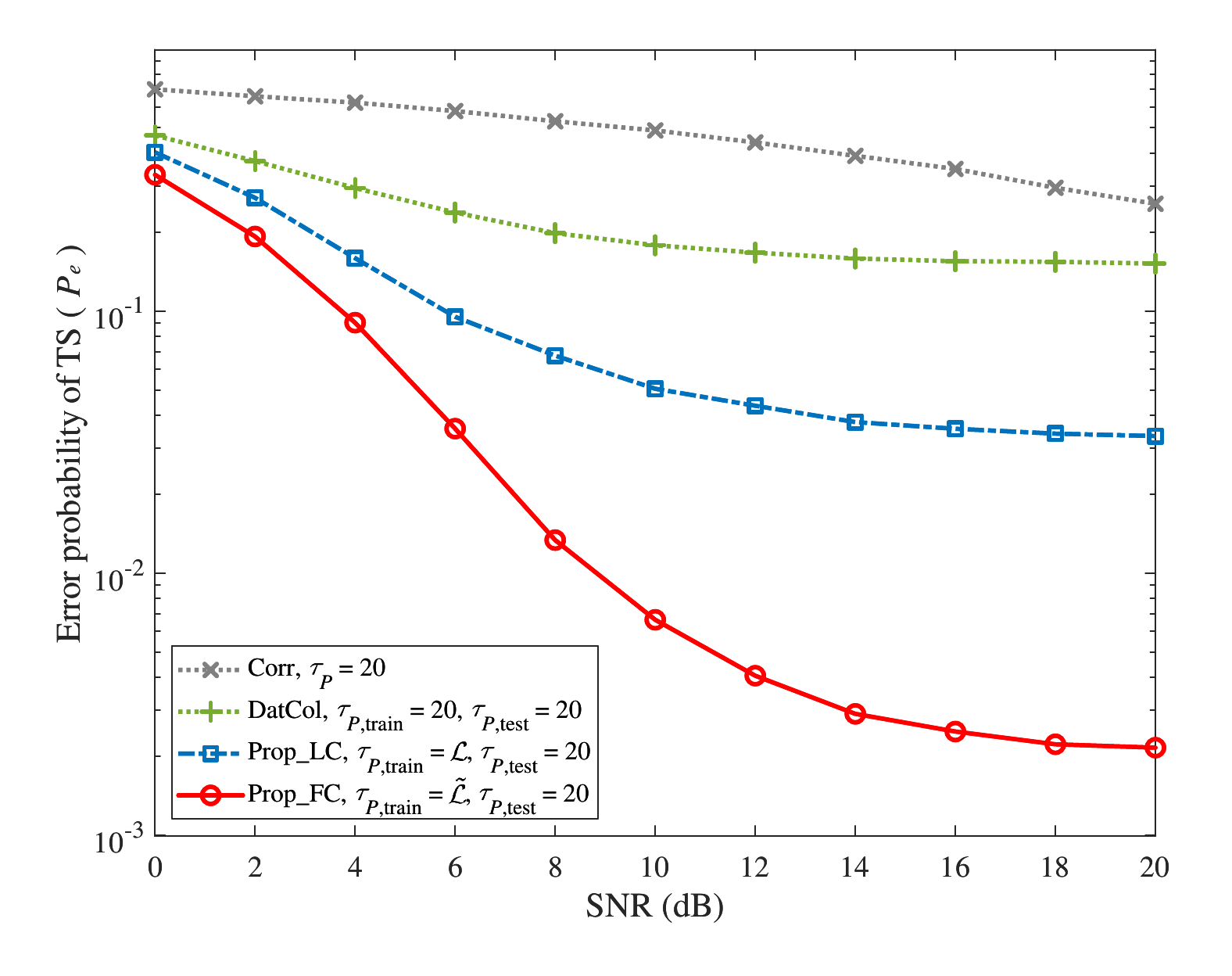}
    }
    \subfigure[Impact of the completeness of training data.]{
    \label{fig_comp2}
    \includegraphics[width=0.5\textwidth]{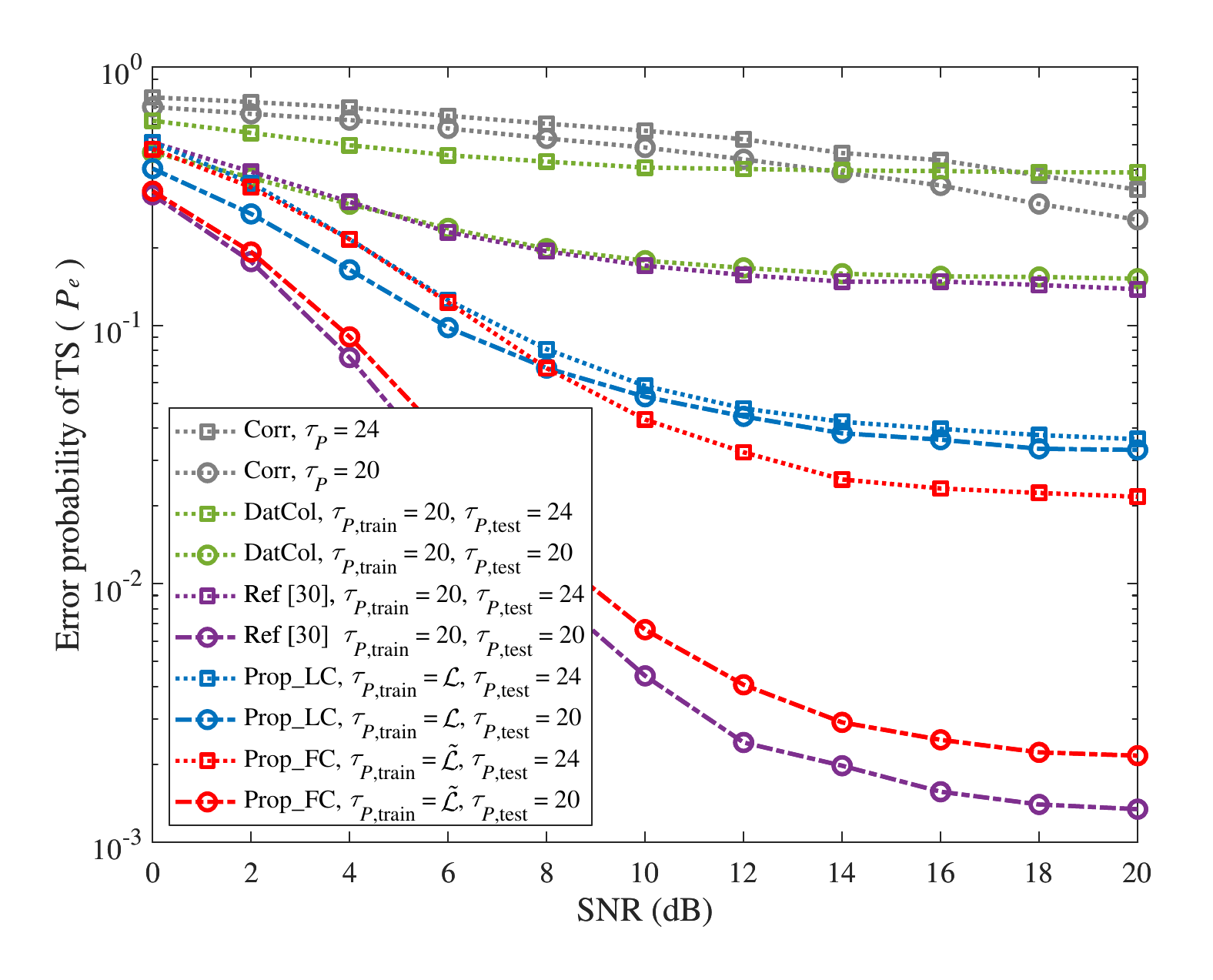}
    }
\caption{The error probability of TS vs. SNRs, where $N=128$, $L_c=32$, and $\eta=0.2$. $\mathcal{L}$ and $\widetilde{\mathcal{L}}$ denote the proposed loose constraint-based strategy and the proposed flexible constraint-based strategy for generating the training data, respectively. $\tau_{P,\textrm{train}}$ and $\tau_{P,\textrm{test}}$ stand for the training maximum multi-path delay and the testing maximum multi-path delay, respectively.}
\label{fig:Comp}
\end{figure}
\begin{table}[t]
\caption{ Storage Consumption and Bandwidth Occupation vs. Different Strategies.}\label{Tab:CC0}
\centering
\renewcommand\arraystretch{1.25}
\begin{tabular}{|c||c||c|}
\hline
{Strategy}  & {\makecell[c]{Storage\\Consumption (MB)}} & \makecell[c]{Bandwidth\\Occupation (sec)}\\ \hline
``DatCol''  & $633.5277$ & $50.6822$\\ \hline
``Prop\_LC'' and ``Prop\_FC'' & $469.7252$ & $0$ \\ \hline
\end{tabular}
\end{table}
\subsection{Effectiveness Analysis}\label{VB}
In Fig.~\ref{fig_comp1} and Fig.~\ref{fig_comp2}, the accuracy and completeness of learning labels are validated, respectively. Additionally, the storage consumption and bandwidth occupation of different strategies for obtaining training data are given in Table \ref{Tab:CC0}.

From Fig.~\ref{fig_comp1}, the error probability of ``DatCol'' is higher than those of the proposed ``Prop\_LC'' and ``Prop\_FC''. This is due the fact that the collected training data set includes the incorrect learning labels according to \eqref{EQ:Pr2}. Therefore, the existed strategy of training data collection (i.e., ``DatCol'') inevitably deteriorates the performance of ELM-based TS network, increasing the error probability of TS. By contrast, the proposed strategies of ``Prop\_LC'' and ``Prop\_FC'' achieve smaller error probabilities of TS than ``DatCol''. This indicates that the proposed strategy for training data generation avoids incorrectly estimating learning labels, due to the good controllability of computer-aided data generation.

To present the merits of the proposed TS methods in improving the TS correctness when changing wireless scenarios, Fig.~\ref{fig_comp2} depicts the error probability of TS by changing the testing maximum multi-path delay.
From Fig~\ref{fig_comp2}, when $\tau_{P,\textrm{test}}>\tau_{P,\textrm{train}}$, the ELM-based TS network trained with the strategies of ``DatCol'' and ``Ref \cite{ref:ELM-labelTS}'' reach higher error probabilities than those of ``Prop\_LC'' and ``Prop\_FC'', due to the impacts of the impaired completeness of training data. Especially for ``Ref \cite{ref:ELM-labelTS}'' in the whole SNR region, its error probabilities significantly increase when $\tau_{P,\textrm{test}}>\tau_{P,\textrm{train}}$. For example, when $\tau_{P,\textrm{test}}=24$, the error probability of ``DatCol'' is higher than 0.25 for each given value of SNR, and the error probabilities of ``Ref \cite{ref:ELM-labelTS}'' are higher than 0.15 for the relatively high SNRs. As a result, the ELM-based TS networks trained with the existed strategies of ``DatCol'' and ``Ref \cite{ref:ELM-labelTS}'' hinder themselves to the practical application for the impaired completeness of training data. By contrast, after training the TS network, the error probabilities of ``Prop\_LC'' and ``Prop\_FC'' are much smaller than those of ``DatCol'' and ``Ref \cite{ref:ELM-labelTS}'' when $\tau_{P,\textrm{test}}=24$, for all given SNRs. This reflects that the proposed strategies of ``Prop\_LC'' and ``Prop\_FC'' can generate more complete training data.
This is because ``Prop\_LC'' and ``Prop\_FC'' utilize the channel models with randomly generated multi-path delays (i.e., $\tau_{P,\textrm{train}}=\mathcal{L}$ and $\mathcal{L}\sim(16,26)$) for model training, the diverse $\mathcal{L}$ generates more power delay profiles (PDPs) to increase the diversity of training data for increasing the completeness of training data.
Then, a good completeness of training data can be obtained to facilitate the performance of ELM-based TS networks, reducing the error probability of TS.
For example, when $\mathrm{SNR}_p\ge10$dB and $\tau_{P,\textrm{test}}=24$, the error probabilities of ``Prop\_LC'' and ``Prop\_FC'' are lower than $0.0583$.
Therefore, compared with the existed strategies of ``DatCol'' and ``Ref \cite{ref:ELM-labelTS}'' for obtaining training data, the proposed computer-aided strategies for generating training data (i.e., ``Prop\_LC'' and ``Prop\_FC'') can obtain more complete training data set, providing improvement in the reduction of TS errors against the parameter of $\tau_{P,\textrm{test}}$.

According to \eqref{EQ:Pr2}, the estimated learning labels are critically dependent on the error probability of TS, during the training data collection. By assuming the training data collected under $\mathrm{SNR}_p=20$dB as shown in Fig.~\ref{fig:Comp}, the error probability of TS equals about 0.257 so that the accuracy of estimated learning labels is about $0.743$. To collect $10^5$ effective training samples (i.e., $10^5$ correct learning labels), at least $1.35\times10^5$ received samples are required. Since the collected samples need to be transmitted by the device of data collection to the Central Processing Unit, there are storage and bandwidth occupation. From Table~\ref{Tab:CC0}, we can see that, the storage cost of the proposed strategies (i.e., ``Prop\_LC'' and ``Prop\_FC'') are less than that of ``DatCol'', since the method of data generation is known to the transceiver. That is, the training data can be timely generated and released, leading to less consumption of storage compared with ``DatCol''. Furthermore, there are no incorrect learning labels and the data transmission (i.e., bandwidth occupation) is avoided. If the transmission bandwidth is given by $100$MHz, it could be obtained that ``DatCol'' requires 50.6822 sec for transmitting the collected samples, as shown in Table~\ref{Tab:CC0}. By contrast, ``Prop\_LC'' and ``Prop\_FC'' avoid bandwidth occupation for transmitting training samples during the wireless transmission, i.e., the bandwidth occupation in sec equals to 0.
\emph{This is because, the training data is generated at local receiver, which does not occupy the transmission bandwidth.}
As a result, compared with ``DatCol'', the proposed ``Prop\_LC'' and ``Prop\_FC'' significantly improve the storage consumption and bandwidth occupation.

In summary, compared with ``DatCol'', the error probability of TS reveals that the proposed strategies of ``Prop\_LC'' and ``Prop\_FC'' can avoid the incorrect learning labels and improve the completeness of training data. In addition, compared with ``DatCol'' in  Table~\ref{Tab:CC0}, the proposed strategies of ``Prop\_LC'' and ``Prop\_FC'' possess less storage consumption and bandwidth occupation, which are friendly to practical implementation.

\subsection{Generalization Analysis}\label{VC}
\begin{figure}[t]
\centering
\includegraphics[width=0.5\textwidth]{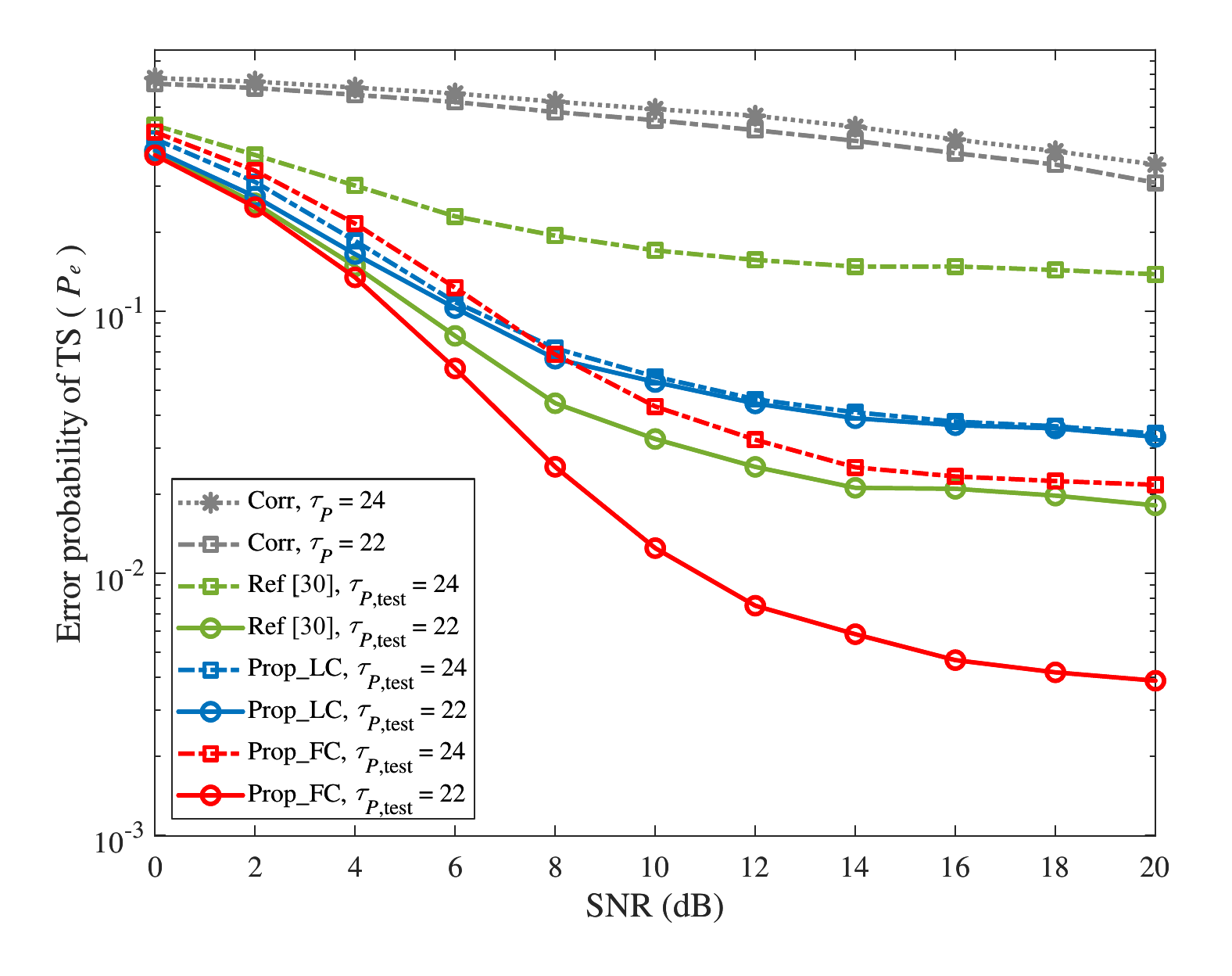}\\
\caption{The generalization analysis with respect to the maximum multi-path delay, where $N=128$, $L_c=32$, $\eta=0.2$, and different testing values of $\tau_{P,\textrm{test}}$ (i.e., $\tau_{P,\textrm{test}}=22$ and $\tau_{P,\textrm{test}}=24$) are considered.}\label{Fig:Gen_Val_tau}
\end{figure}
The more complete the training data is, the stronger the statistical efficiency of the neural network will be. Hence, we utilize the error probability of TS  to validate the generation performance of the designed TS learner (i.e., the ELM-based TS network with training strategies of ``Prop\_LC'' and ``Prop\_FC''). In addition, we do not discuss the generalization performance of ``DatCol'', since the TS performance of ``DatCol'' is mainly affected by incorrect learning labels in the training data collection.
Except for the parameter that needs to discuss, other basic parameters remain the same as those
adopted in Section IV-B.

From Fig.~\ref{Fig:Gen_Val_tau}, for each given value of $\tau_{P,\textrm{test}}$, the error probability of ``Prop\_FC'' is less than those of ``Corr'' and ``Ref \cite{ref:ELM-labelTS}'' in the whole SNR region. This is due to the fact that the proposed strategy of ``Prop\_FC'' can generate more complete training data, which improves the generalization performance against the impact of $\tau_{P,\textrm{test}}$. As for $\tau_{P,\textrm{test}}=22$, ``Ref \cite{ref:ELM-labelTS}'' reaches a lower error probability of TS than ``Prop\_LC''. Even so, with $\tau_{P,\textrm{test}}$ increases to $24$, the error probability of ``Ref \cite{ref:ELM-labelTS}'' is significantly higher than that of ``Prop\_LC'', for the relatively high SNRs. As a result, compared with the existed strategy of ``Ref \cite{ref:ELM-labelTS}'', the proposed strategies of ``Prop\_LC'' and ``Prop\_FC'' are able to deal with the varied maximum multi-path delays.

\subsection{Robustness Analysis}\label{VD}
\begin{figure}[t]
\centering
\includegraphics[width=0.5\textwidth]{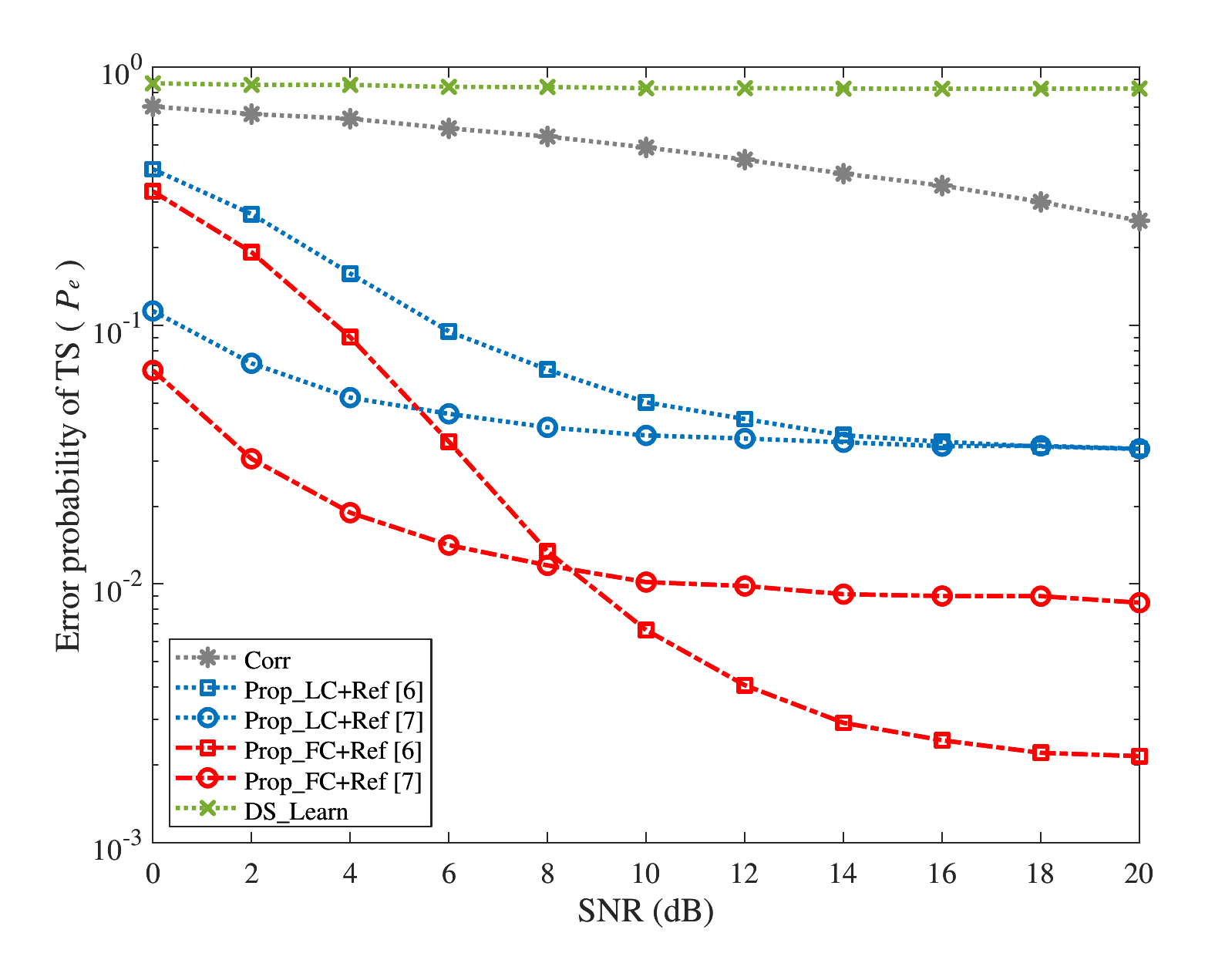}\\
\caption{The Error probability of TS vs. SNRs, where $N=128$, $L_c=32$, $\eta=0.2$, $\tau_P=20$, and different TS schemes are used for feature extraction.}\label{Fig:Rob_Val_L}
\end{figure}

\begin{figure}[t]
\centering
\includegraphics[width=0.5\textwidth]{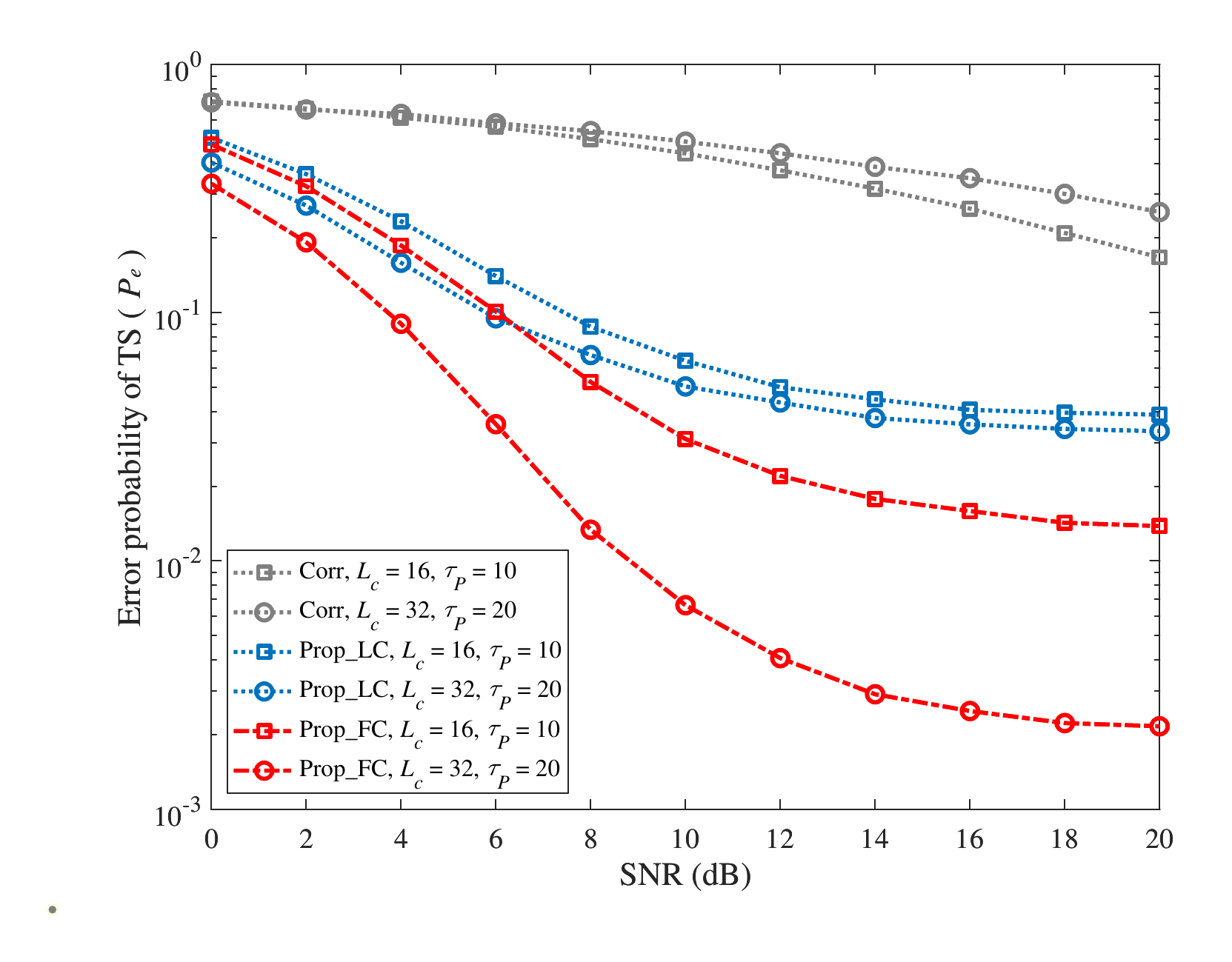}\\
\caption{The error probability of TS vs. SNRs, where $N=128$, $\eta=0.2$, and different values of CP length (i.e., $L_c=16$ and $L_c=32$) are considered. Correspondingly, when $L_c=16$, $\mathcal{L}$ is set as 13 and $\widetilde{\mathcal{L}}$ satisfies that $\widetilde{\mathcal{L}}\sim\mathcal{U}(8,13)$. As for $L_c=32$, $\mathcal{L}$ is set as 26 and $\widetilde{\mathcal{L}}$ satisfies that $\widetilde{\mathcal{L}}\sim\mathcal{U}(16,26)$.
}\label{Fig:Rob_Val_L_c}
\end{figure}

This subsection presents the robustness analysis of the proposed TS scheme against the changed parameters such as the length of CP (i.e., $L_c$), the size of OFDM symbol (i.e., $N$), and the decay factor (i.e., $\eta$), etc. In addition, different TS schemes (e.g., \cite{ref:SC,ref:Minn}) are adopted as the feature extraction, which aims to reveal the robustness against the feature extraction. The error probabilities of the proposed TS scheme are plotted from Fig. \ref{Fig:Rob_Val_L} to Fig. \ref{Fig:Rob_Val_Eta}. For the sake of clarity, ``Ref \cite{ref:SC}'' and ``Ref \cite{ref:Minn}'' denote that the TS schemes mentioned in \cite{ref:SC} and \cite{ref:Minn} are utilized for the feature extraction, respectively. The ELM-based TS network which directly learns the received samples is denoted as ``DS\_Learn'', i.e., leveraging the received samples as the ELM input without the procedure of feature extraction. In the following simulations, unless otherwise specified, parameters remain the same as those mentioned in Section \ref{VB}.
\subsubsection{Robustness Against the TS Scheme for Feature Extraction}
Fig.~\ref{Fig:Rob_Val_L} illustrates the error probability of TS to present the robustness against the impact of feature extraction. From Fig.~\ref{Fig:Rob_Val_L}, it could be obtained that, the error probability of ``DS\_Learn'' is higher than that of ``Corr'', even for the relatively high SNRs. This reflects that the ELM-based TS network cannot work well without the process of feature extraction. For each method of feature extraction, the error probabilities of ``Prop\_LC'' and ``Prop\_FC'' are smaller than that of ``Corr''. This is due to the fact that the TS schemes as feature extraction successfully extract the hidden state information about STOs, and thus improve the correctness of estimating STOs. In addition, the ELM-based TS networks with different feature extractions exhibit different performance improvements in reducing the error probability of TS. Several reasons may be used to explain this result. {1)} The TS scheme of \cite{ref:Minn} is vulnerable to the impact of channel fading. {2)} The TS scheme of \cite{ref:Minn} provides a sharper correlation peak compared with that of \cite{ref:SC}, but the timing index of the correlation peak is largely independent of the ISI-free region. {3)} The TS scheme in \cite{ref:SC} has robust features highlighting the ISI-free region compared with that in \cite{ref:Minn}. Therefore, the ELM-based TS network assisted by the TS scheme of \cite{ref:SC} outperforms that assisted by \cite{ref:Minn}, for the relatively high SNRs. To sum up, compared with ``Corr'' and ``DS\_Learn'', the ELM-based TS network with feature extraction can reach significantly lower error probabilities of TS.

\subsubsection{Robustness Against $L_c$}
To reveal the robustness of the proposed TS scheme against the impact of $L_c$, Fig.~\ref{Fig:Rob_Val_L_c} illustrates the error probability of TS by changing $L_c$. From Fig.~\ref{Fig:Rob_Val_L_c}, for each given value of $L_c$, the error probabilities of ``Prop\_LC'' and ``Prop\_FC'' are smaller than that of ``Corr'' in the whole SNR region. It is worth noting that, by decreasing $L_c$, the error probabilities of ``Prop\_LC'' and ``Prop\_FC'' are enlarged. Especially, ``Prop\_FC'' reaches a smaller error probability than ``Prop\_LC''. The reason is that ``Prop\_FC'' owns a more complete profile of CIRs, which facilitates the statistical efficiency of the TS network. Although the ISI-free region is shortened by narrowing the length of CP, the proposed  ``Prop\_LC'' and ``Prop\_FC'' still obtain a robust improvement in reducing the error probability of TS.
\begin{figure}[t]
\centering
\includegraphics[width=0.5\textwidth]{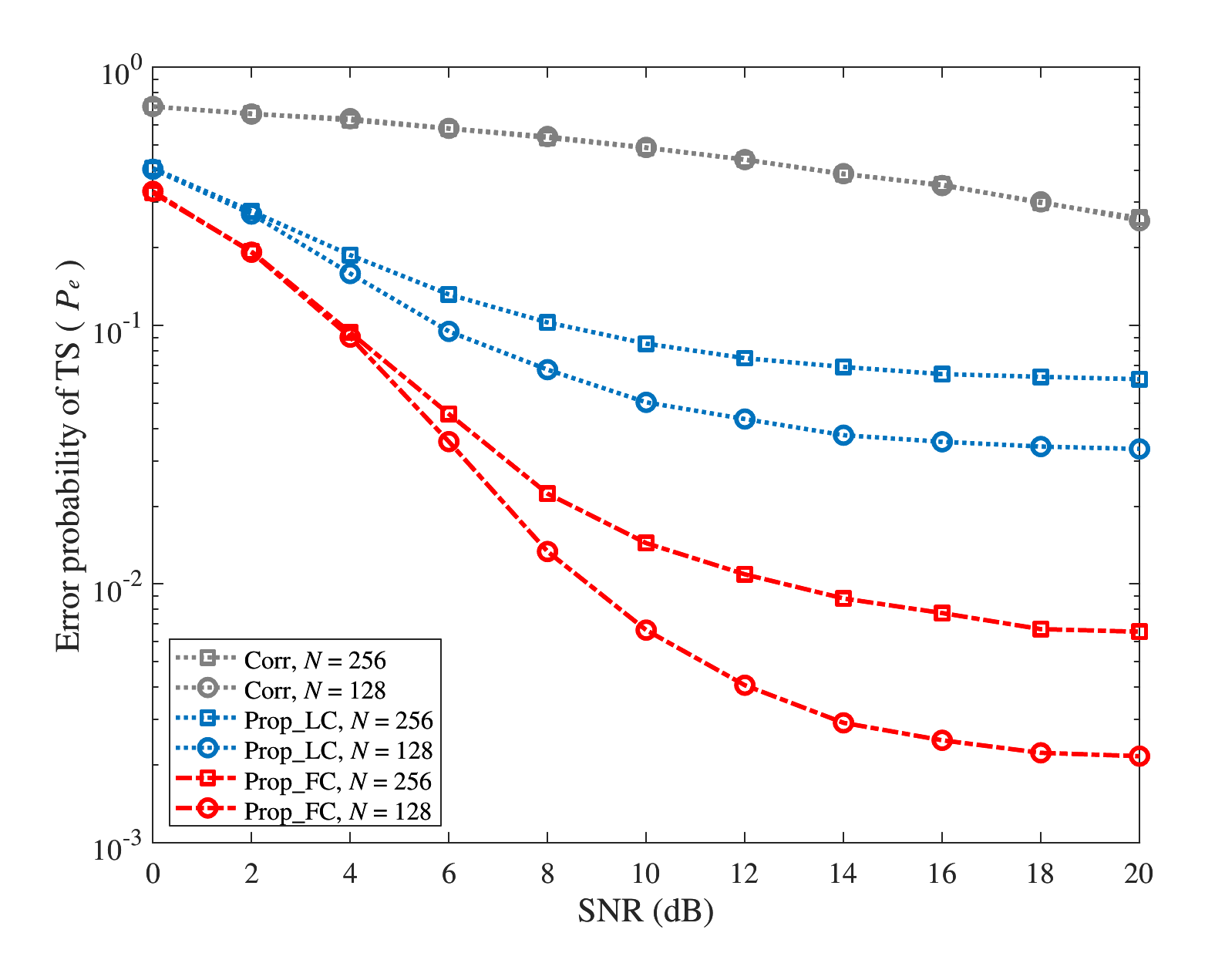}\\
\caption{The error probability of TS vs. SNRs, where $\tau_{P}=20$, $L_c=32$, $\eta=0.2$, and different values of symbol size (i.e., $N=128$ and $N=256$) are considered.
}\label{Fig:Rob_Val_N}
\end{figure}
\subsubsection{Robustness Against $N$}
Fig. \ref{Fig:Rob_Val_N} depicts the error probability of TS against the impacts of varying $N$. From Fig. \ref{Fig:Rob_Val_N}, for each given value of $N$, the error probability of ``Corr'' is higher than those of ``Prop\_LC'' and ``Prop\_FC'' in the whole SNR region.
Among the proposed strategies of ``Prop\_LC'' and ``Prop\_FC'', ``Prop\_FC'' reveals a remarkable improvement in the reduction of error provability relative to ``Prop\_LC'', for each given $N$.
This is due to the fact that, ``Prop\_FC'' has the ability to improve the correlation between the training labels and the ground truth labels, and thus provides improvements in reducing error probability of TS.
For example, when $\mathrm{SNR}_p=12$dB and $N=256$, the error probability of ``Prop\_LC'' is around $0.084$, while it achieves 0.014 for ``Prop\_FC''.
Although the error probability of TS increases with the increase of $N$, the proposed ``Prop\_LC'' and ``Prop\_FC''  still outperform ``Corr''.
As a result, the proposed strategies of ``Prop\_LC'' and ``Prop\_FC'' achieve robust improvements in the reduction of  TS errors against the impact of $N$.
\begin{figure}[t]
\centering
\includegraphics[width=0.5\textwidth]{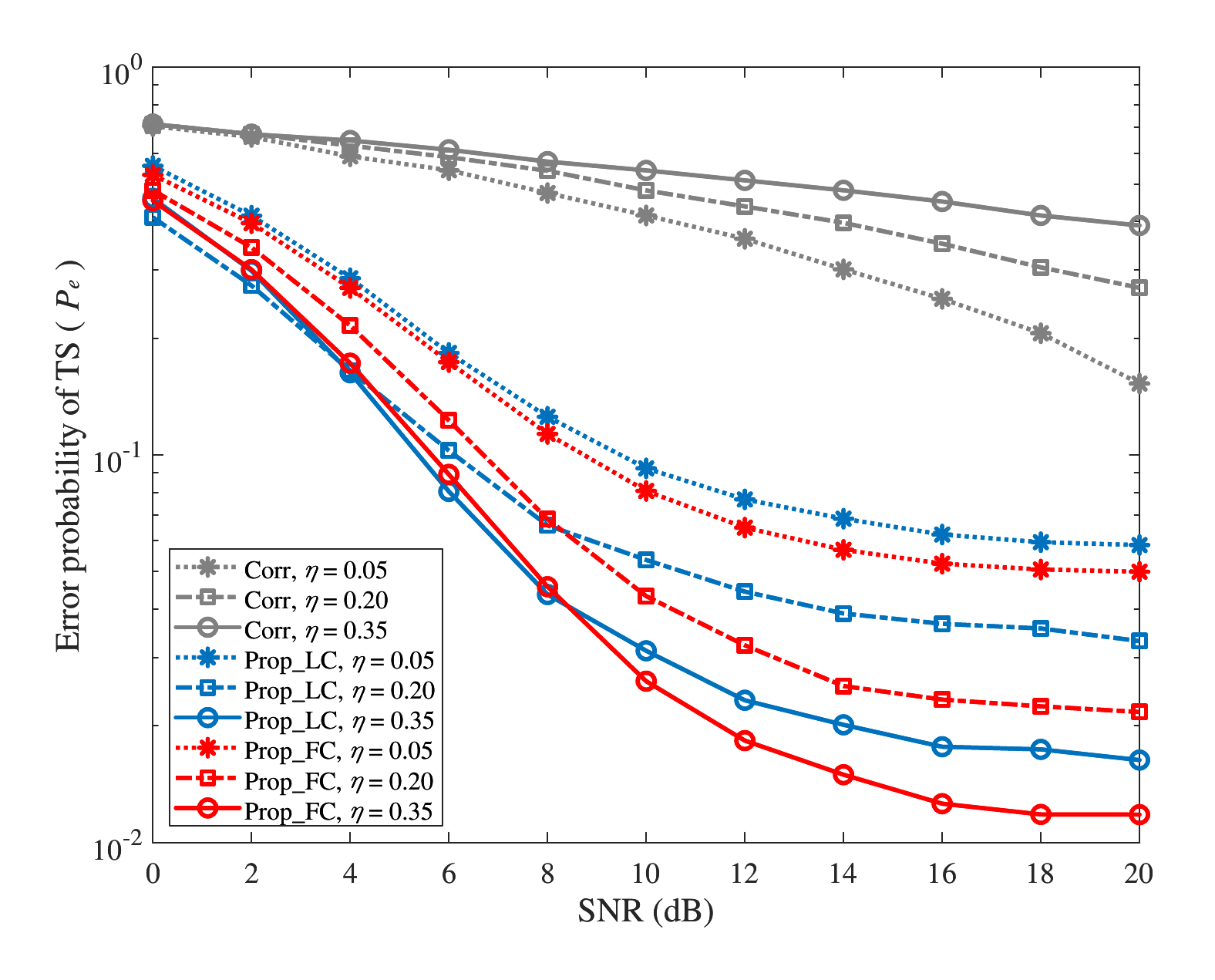}\\
\caption{The error probability of TS vs. SNRs, where $\tau_{P}=20$, $L_c=32$, $N=128$, and different values of decay factor (i.e., $\eta=0.05$, $\eta=0.20$, and $\eta=0.35$) are considered.
}\label{Fig:Rob_Val_Eta}
\end{figure}
\subsubsection{Robustness Against $\eta$}
To reflect the robustness of the proposed TS scheme against the impact of $\eta$, Fig.~\ref{Fig:Rob_Val_Eta} depicts the error probability of TS where $\eta$ varies from 0.05 to 0.35 with the interval being 0.15.
In Fig.~\ref{Fig:Rob_Val_Eta}, the error probability of ``Corr'' decreases with the decrease of $\eta$. Because the correlation platform/peaks obtained by ``Corr'' have been enhanced by the large value of $\eta$, and so that timing instants of the maximum peak appear in advance to the ISI region of CP.
For each given value of $\eta$ in Fig.~\ref{Fig:Rob_Val_Eta}, the error probability of ``Corr'' is higher than those of ``Prop\_LC'' and ``Prop\_FC'' in the whole SNR region. For the relatively high SNRs, the error provability of ``Prop\_FC'' is smaller than that of ``Prop\_LC''. This indicates the superiority of ``Prop\_FC'' in reducing the TS errors. Also, it could be observed that, the error probabilities of ``Prop\_LC'' and ``Prop\_FC'' decrease with the increase of $\eta$. This is likely due to the enlarged $\eta$, which strengthens the power of the first propagation path. The stronger the first propagation path is, the stronger timing features will be, such that improves the learning efficiency of ELM-based TS network. In addition, although the differences of these error probabilities are not obvious to the varied $\eta$, ELM-based TS networks with the proposed strategies of ``Prop\_LC'' and ``Prop\_FC'' still outperform ``Corr'' in the aspect of reducing TS errors.
In a word, compared with ``Corr'', the proposed strategies of ``Prop\_LC'' and ``Prop\_FC'' can provide the robust and effective improvement in reducing the error probability of TS.
\section{Conclusion}\label{S:V}
In this paper, we propose two computer-aided strategies to generate the training data, and construct an ELM-based TS network for the OFDM systems. Compared with training data collected from the realistic systems, the desired training data could be timely generated by our proposed strategies at the local device, and this avoids the occupations of storage space and transmission bandwidth during the training data collection. Furthermore, the proposed computer-aided training strategies increase the completeness of training data against the maximum multi-path delay, therefore improving statistical efficiency of the ELM-based TS network. By simulations, numerical results exhibit that the proposed ELM-based TS scheme is superior in reducing the error probability of TS and possesses a good generalization performance against the maximum multi-path delay.

\section{Acknowledgement}
This work is supported in part by the Sichuan Science and Technology Program: (Grant-No. 2021JDRC0003, 2023YFG0316, 2021YFG0064), the Demonstration Project of Chengdu Major Science and Technology Application (Grant No. 2020-YF09- 00048-SN),the Special Funds of Industry Development of Sichuan Province (Grant No. zyf-2018-056), and the Industry University Research Innovation Fund of China University: (Grant No. 2021ITA10016).

\ifCLASSOPTIONcaptionsoff
\newpage
\fi


%\nocite{*}
%\bibliographystyle{ieeetran}
%\bibliography{ref}

\end{document}